\documentclass[a4paper]{spie_margins}

\usepackage{amsmath,amsfonts,amssymb}
\usepackage{graphicx}
\usepackage[colorlinks=true, allcolors=blue]{hyperref}

\usepackage{textcomp}    
\usepackage{xcolor}      
\usepackage{aas_macros}  

\graphicspath{{./images/}}  

\title{The Gravitational-wave Optical Transient Observer (GOTO)}

\author[a]{\mbox{Martin J. Dyer}}
\author[b]{\mbox{Kendall Ackley}}
\author[b]{\mbox{Joseph Lyman}}
\author[b]{\mbox{Krzysztof Ulaczyk}}
\author[b]{\mbox{Danny Steeghs}}
\author[c]{\mbox{Duncan K. Galloway}}
\author[a,d]{\mbox{Vik S. Dhillon}}
\author[e]{\mbox{Paul O'Brien}}
\author[f]{\mbox{Gavin Ramsay}}
\author[g]{\mbox{Kanthanakorn Noysena}}
\author[h]{\mbox{Rubina Kotak}}
\author[i]{\mbox{Rene Breton}}
\author[j]{\mbox{Laura Nuttall}}
\author[d]{\mbox{Enric Pallé}}
\author[b]{\mbox{Don Pollacco}}
\author[ ]{\mbox{the GOTO Collaboration}}

\affil[a]{Department of Physics and Astronomy, University of Sheffield, Sheffield S3 7RH, UK}
\affil[b]{Department of Physics, University of Warwick, Coventry CV4 7AL, UK}
\affil[c]{School of Physics \& Astronomy, Monash University, Clayton VIC 3800, Australia}
\affil[d]{Instituto de Astrofísica de Canarias, E-38205 La Laguna, Tenerife, Spain}
\affil[e]{School of Physics \& Astronomy, University of Leicester, University Road, Leicester LE1 7RH, UK}
\affil[f]{Armagh Observatory \& Planetarium, College Hill, Armagh, BT61 9DG, UK}
\affil[g]{National Astronomical Research Institute of Thailand, 260 Moo 4, T. Donkaew, A. Maerim, Chiangmai, 50180
Thailand}
\affil[h]{Department of Physics \& Astronomy, University of Turku, Vesilinnantie 5, Turku, FI-20014, Finland}
\affil[i]{Jodrell Bank Centre for Astrophysics, Department of Physics and Astronomy, The University of Manchester, Manchester M13 9PL, UK}
\affil[j]{Institute of Cosmology \& Gravitation, University of Portsmouth, Portsmouth PO1 3FX, UK}

\authorinfo{Send correspondence to MJD - email: martin.dyer@sheffield.ac.uk}

\begin{document} 
\maketitle

\begin{abstract}
The Gravitational-wave Optical Transient Observer (GOTO) is a wide-field telescope project focused on detecting optical counterparts to gravitational-wave sources. Each GOTO robotic mount holds eight 40 cm telescopes, giving an overall field of view of 40 square degrees. As of 2022 the first two GOTO mounts have been commissioned at the Roque de los Muchachos Observatory on La Palma, Canary Islands, and construction of the second node with two additional 8-telescope mounts has begin at Siding Spring Observatory in New South Wales, Australia. Once fully operational each GOTO mount will be networked to form a robotic, multi-site observatory, which will survey the entire visible sky every two nights and enable rapid follow-up detections of transient sources.
\end{abstract}


\section{Motivation}
\label{sec:}

\begin{figure}[t]
    \begin{center}
        \includegraphics[width=0.9\linewidth]{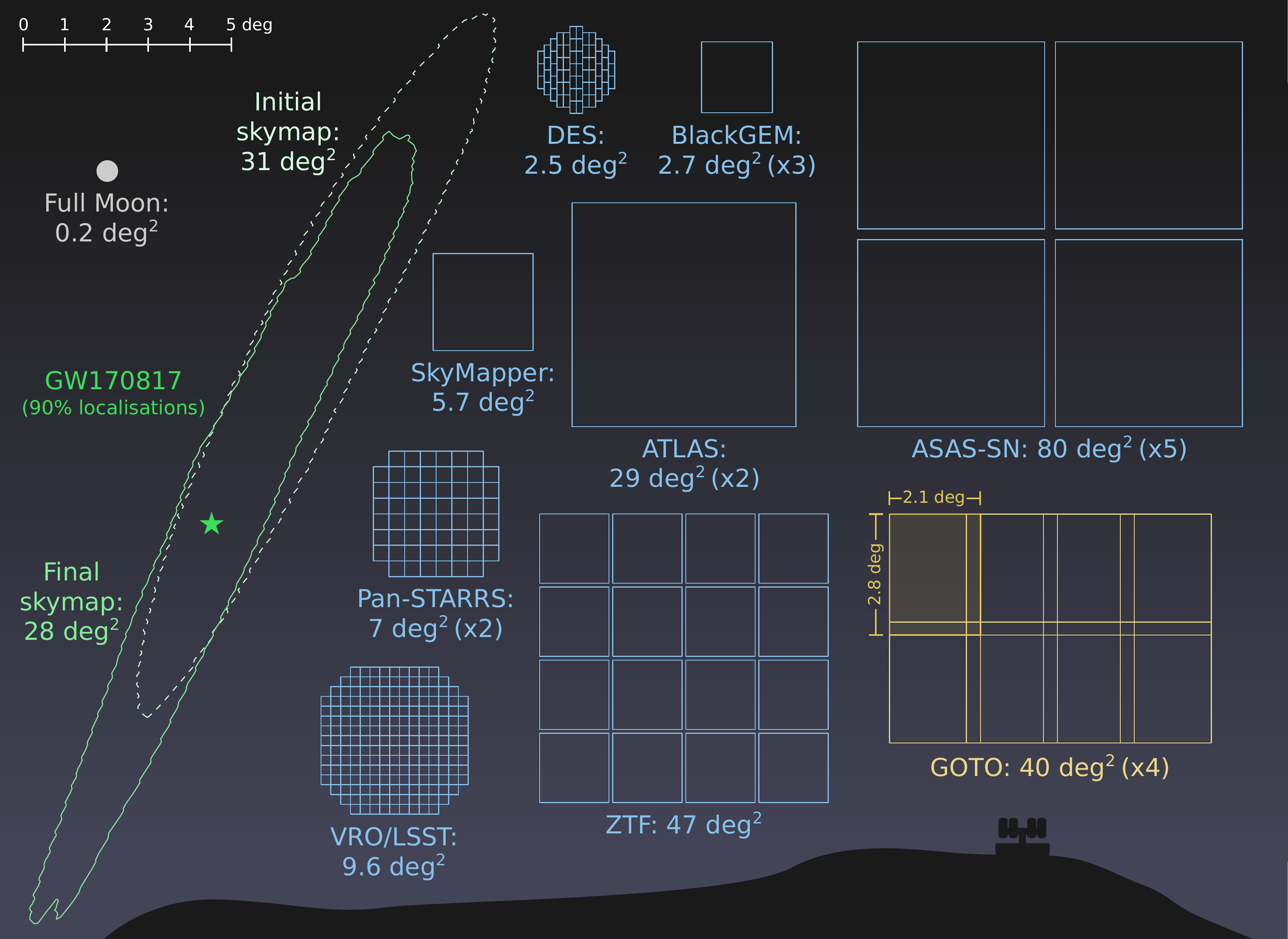}
    \end{center}
    \caption[example]{
        The field of view of a single GOTO mount (in yellow), compared to other selected wide-field telescopes (blue) and the initial and final skymaps for GW170817 (green). The complete 40~square~degree GOTO footprint is formed from the overlapping fields of view of the 8 unit telescopes on the mount, one of which is highlighted.
    }\label{fig:fov}
\end{figure}

Since the first detection of gravitational waves in 2015 \cite{GW150914} only a single counterpart kilonova has been detected, AT~2017gfo relating to GW170817 \cite{GW170817,GW170817_followup}. Several further detections of gravitational waves from binary neutron star inspirals have failed to yield similar detections \cite{GWTC}. A major limiting factor when following up these GW detections is covering the skymaps produced by the detectors quickly enough to detect a kilonova before it fades from view. GW170817 was localised to an area covering approximately 30~square~degrees (shown in Figure~\ref{fig:fov}), allowing telescopes with relatively small fields of view to detect the kilonova using a galaxy-focused strategy\cite{GW170817_followup}. By contrast the second detection of a binary neutron star, GW190425, had an initial skymap that covered over 10,000 square degrees \cite{GW190425}, and in future observing runs it is predicted that the median localisation area will continue to be of the order of thousands of square degrees \cite{DDE}. 

The GOTO project is an international collaboration with the purpose of constructing a network of wide-field, robotic telescopes focused on following-up gravitational wave detections. In this paper we will report on the current status of the project and planned expansion in advance of the O4 gravitational-wave observing run due to begin in 2023.

\newpage

\section{Hardware}
\label{sec:hardware}

In order to cover a large field of view in a cost-effective manner, each GOTO telescope consists of a fast slewing, robotic mount on which is mounted an array of eight ``unit telescopes'' (UTs). Each UT is a $D=40$~cm Wynne–Riccardi OTA, fitted with a focuser, 5-slot filter wheel (with Baader $LRGBC$ filters) and an FLI ML50100 camera with a 50 megapixel CCD detector (KAF-50100). Each UT has a field of view of $2.8 \times 2.1$~square~degrees, and when aligned in an array configuration each mount has an overall field of view of ~40 square degrees (with some overlap between UTs for calibration; shown in Figure~\ref{fig:fov}). When observing using the wide (400-700 nm) $L$ filter each UT can reach a depth of 20-21 magnitude in two minutes. The UTs are mounted on a boom-arm on a German equatorial mount, and each mount is situated in an individual Astrohaven clamshell dome. 

\begin{figure}[p]
    \begin{center}
        \includegraphics[width=0.95\linewidth]{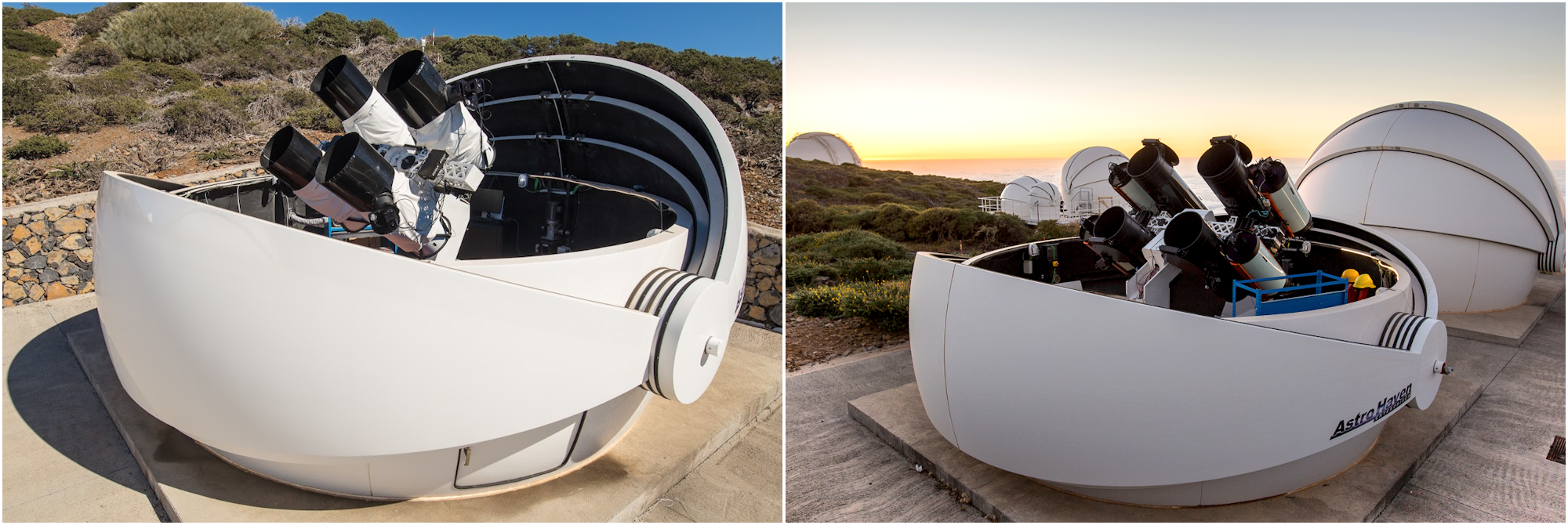}
    \end{center}
    \caption[example]{
        The original GOTO prototype in the dome on La Palma. On the left the initial four unit telescopes are shown in 2019, attached to the prototype mount. On the right in 2020 the prototype unit telescopes have been replaced with the new ASA design, alongside four temporary Celestron RASAs on the outside of the array.
    }\label{fig:photo1}
\end{figure}

\begin{figure}[p]
    \begin{center}
        \includegraphics[width=0.95\linewidth]{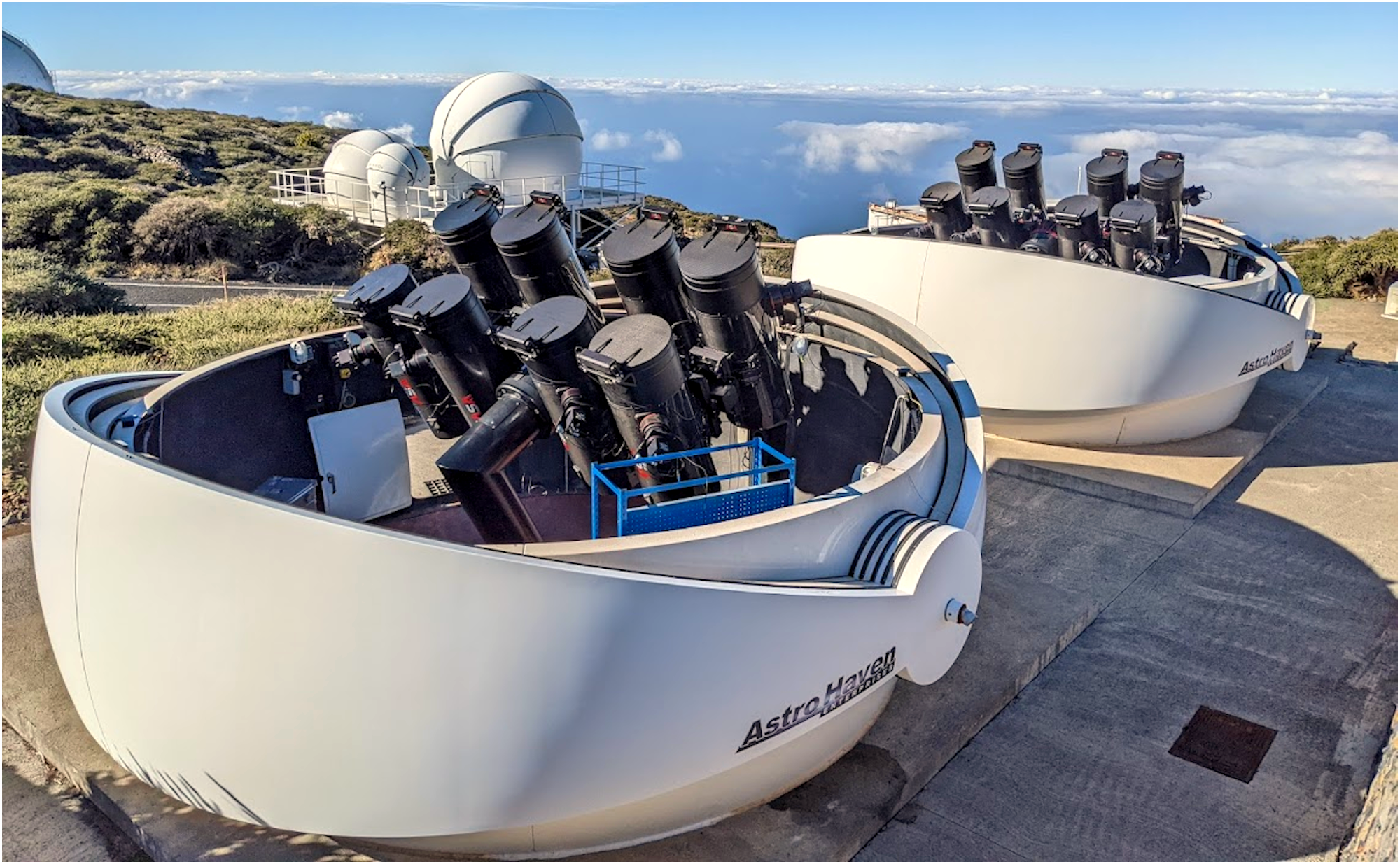}
    \end{center}
    \caption[example]{
        The complete GOTO-North node as of 2021. Each dome hosts a single mount with eight unit telescopes.
    }\label{fig:photo2}
\end{figure}

The GOTO prototype was built by APM Professional Telescopes, with second-generation hardware built by ASA Astrosysteme. The initial GOTO prototype operated as a single mount with four, later eight, unit telescopes (shown in Figure~\ref{fig:photo1}). Following the end of the prototype phase the mount was replaced by a new, upgraded design with a full array of eight unit telescopes. The prototype system saw issues with mount stability and scattered light in the corrector, these were improved upon by upgrading to a direct-drive mount and fully-enclosed tubes \cite{prototype}. A second, identical array with another eight UTs was installed in a second dome at the same site (shown in Figure~\ref{fig:photo2}), completing the northern GOTO node and bringing the possible instantaneous field of view up to 80~square~degrees.

\newpage

\begin{figure}[t]
    \begin{center}
        \includegraphics[width=0.9\linewidth]{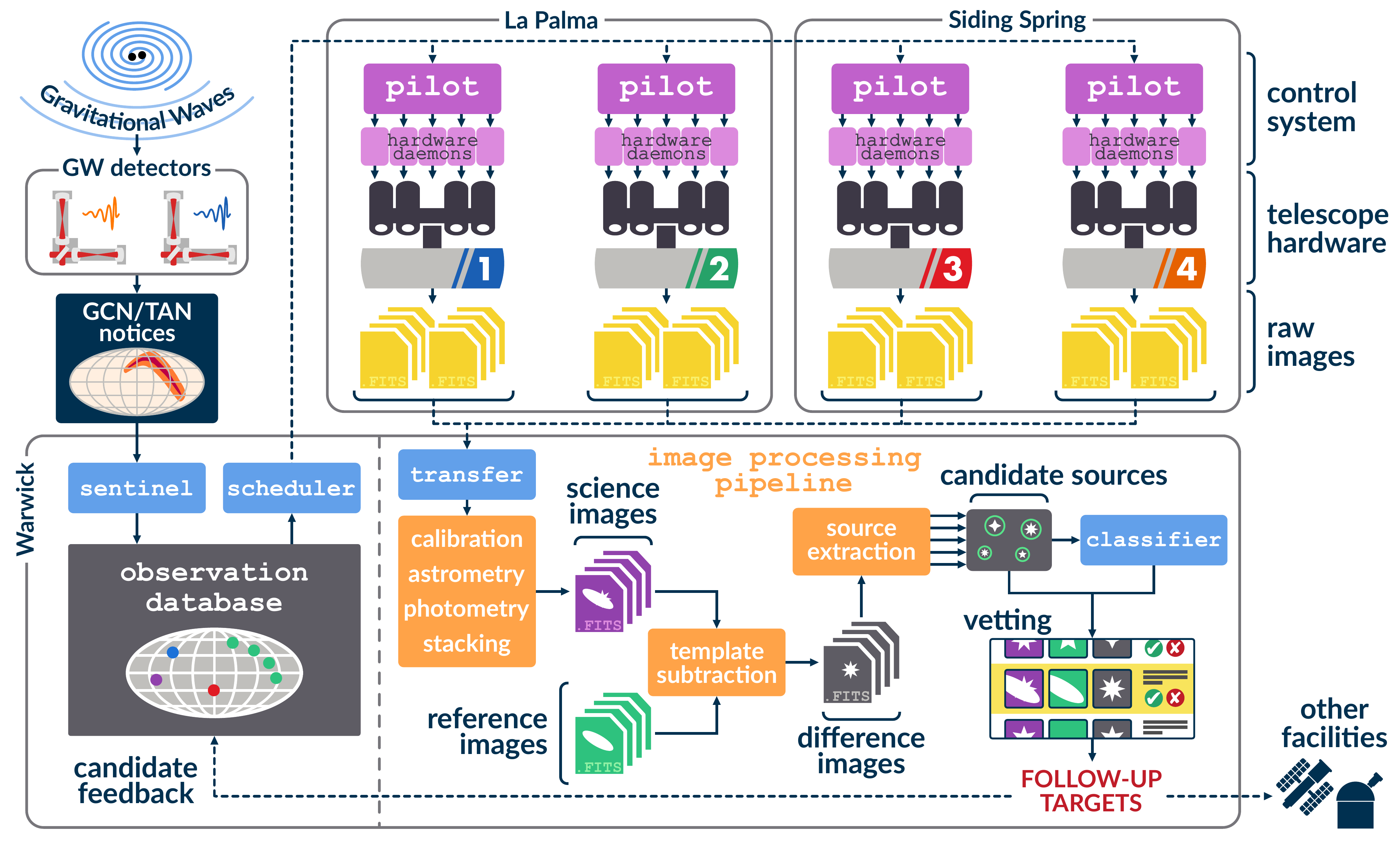}
    \end{center}
    \caption[example]{
        A flow chart visualising the entire GOTO follow-up network. Starting in the top-left, gravitational wave signals are received and processed by the G-TeCS sentinel, with pointings being issued by the scheduler to each of the four telescopes across two sites. The resulting images are transferred back to Warwick where they are processed using reference images to find candidate transient sources.
    }\label{fig:flow}
\end{figure}

\section{Software}
\label{sec:software}

\subsection{Autonomous control}
\label{sec:control}

The GOTO telescopes are designed to operate entirely autonomously, in order to reduce the delay in following-up transient events as far as possible. The GOTO Telescope Control System (G-TeCS) contains the software to process transient events, schedule on-sky pointings and operate the individual telescope hardware units \cite{gtecs2018,gtecs2020}. When in robotic mode each GOTO telescope is controlled by a \textit{pilot} control program, which monitors and sends commands to each piece of hardware through a series of hardware daemons covering each aspect of the telescope (cameras, mount, dome etc...). The pilot operates the telescope each night in place of a human observer, from taking calibration frames in the evening to observing targets until sunrise the next morning. A robust conditions monitoring system ensures that the domes will close in bad weather, allowing the telescopes to operate without human monitoring. Each GOTO telescope has its own pilot and daemons (shown in Figure~\ref{fig:flow}), allowing them to act independently in the case of any failure.

\subsection{Alert processing and scheduling}
\label{sec:scheduling}

In order to build up an all-sky survey of reference images, all observations made by GOTO telescopes are taken aligned to a fixed sky grid \cite{thesis}. Transient alerts from the GCN/TAN network \cite{GCN} are received by the \textit{sentinel} alert monitoring daemon, which processes the alert and determines which grid tiles should be added to the observation database. Based on the valid pointings in the database, the central \textit{scheduler} program calculates the optimal target for each telescope in the network and communicates it to the telescope pilots. The pilots then inform the scheduler when each target is finished, whereupon the scheduler replies with the next highest priority target. The scheduler also continuously recalculates which target in the database has the highest priority every 5 seconds, making it very quick to react to any new transient alert targets added by the sentinel. The pilots regularly check the scheduler while observations are ongoing, meaning any new high-priority targets can be triggered as soon as possible.

\newpage

This system of a central scheduler issuing orders to multiple independent mounts allows for a reactive and flexible scheduling system. It is envisioned that the final multi-site scheduler will be able to alter follow-up strategies to best make use of the entire network, for instance spreading out coverage for wide search regions or focusing on simultaneous observations of single fields in different filters for small ones \cite{thesis}.

\subsection{Image processing}
\label{sec:pipeline}

\begin{figure}[t]
    \begin{center}
        \includegraphics[width=0.9\linewidth]{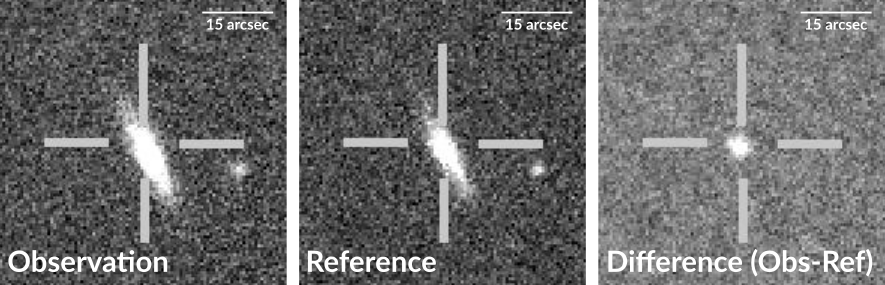}
    \end{center}
    \caption[example]{
        An example of the difference imaging process: SN 2019bpc (aka ASASSN-19ew/Gaia19axa/ZTF19aalvaxt), a Type Ia supernova independently detected by the GOTO prototype pipeline \cite{prototype}.
    }\label{fig:diff}
\end{figure}

Frames from each of the GOTO telescopes are transferred back to the central processing servers located at Warwick University, UK, using a dedicated fibre link. Standard image calibration tasks (bias, dark,
and flat-field corrections) are carried out on each frame, along with source detection using \texttt{SExtractor} \cite{sextractor} and an astrometric solution is then found using
\texttt{astrometry.net} \cite{astrometrydotnet}. When the pilot is observing it takes multiple images of each tile positionto from exposure sets, typically 3 or 4 exposures each of 30--120~s. Once each image within a set is processed they are aligned and median-combined, and this combined frame is then reprocessed through the same source detection, astrometry and photometry steps to produce the final science image for that observation. In addition to adding extra depth to the final stack, taking multiple exposures per pointing allows the detection of moving objects in-between frames.

In order to detect any transient sources that have recently appeared in the new images, a reference image for the same position on the sky is subtracted from each new science image using \texttt{HOTPANTS} \cite{hotpants}. These reference images are built up by the all-sky survey, which aims to ensure there is a recent reference for every visible grid tile position. The resulting difference images are then passed through \texttt{SExtractor} again, which produces a list of transient candidates (an example is shown in Figure~\ref{fig:diff}).

\subsection{Source identification}
\label{sec:sources}

The difference imaging process inevitably produces multiple subtraction artefacts that need to be filtered out of the list of potential transient source candidates. Various methods are employed to rank each detection on a "real-bogus" scale, with a convolutional neural network trained on archival data producing scores that can help filter out spurious detections \cite{Killestein2021}.

Once each detection has a score assigned it is published to a web interface called the GOTO Marshall, where collaboration members can review and flag any promising detections. Contextual information is also included, such as image-stamps, light-curve plots and cross-matches with astronomical catalogues and minor planet sources. Recurrent neural networks have also been developed to provide source classification based on their light-curves \cite{Burhanudin2021}. Work is ongoing to make a publicly-accessible version of the Marshall, utilising the successful Zooniverse platform \cite{galaxyzoo}.

\newpage

\section{Deployment and performance}
\label{sec:deployment}

\subsection{Prototype phase (2017-2021)}
\label{sec:phase1}
The GOTO prototype was deployed at the Observatorio del Roque de los Muchachos on La Palma in the Canary Islands in the summer of 2017 \cite{goto2020,prototype}. The prototype consisted of a single mount and four first-generation unit telescopes, located in the southern of the two newly-constructed domes. After an extended commissioning period the prototype began taking regular observations in February 2019, and covered the entire LIGO-Virgo O3 observing run from April 2019 until its suspension in March 2020. During the first half of the observing run (O3a) GOTO covered tiles from 32 GW triggers \cite{Gompertz2020}, notably contributing to a multi-facility follow-up effort of GW190814 as part of the ENGRAVE collaboration \cite{G190814_ENGRAVE} GOTO covered a mean area of 732~square~degrees per alert, with the shortest time between an alert being received and the exposures beginning being 28~s. During O3 no further kilonova candidates were found, by GOTO or any other facility \cite{GWTC}.

During the prototype period GOTO carried out an all-sky survey to provide reference images, observing each of the approximately 2200 visible grid tiles an average of 20 times over the 18-month period. The prototype also followed up 77 gamma-ray burst (GRB) triggers, detecting four optical counterparts including that for GRB 171205A \cite{Mong2020,GRB171205A}, monitored several AM CVn systems \cite{Duffy2021}, and carried out various other scientific and commissioning tests \cite{prototype}.

The GOTO prototype provided a good demonstration of how the multi-telescope array design could meet the scientific goals of the project. Lessons learnt from the prototype stage were taken into account in improvements to the second generation hardware, along with further developments and upgrades to the control, scheduling and image processing software.

\subsection{GOTO-North (2021+)}
\label{sec:phase2}
Starting in 2020 the GOTO prototype was gradually decommissioned and replaced by the new, second-generation telescopes. The first set of four new unit telescopes was initially tested on the prototype mount in late 2020 (shown on the right in Figure~\ref{fig:photo1}), before the first of the new mounts was deployed in the second dome in summer 2021. The prototype mount was finally removed and replaced with the second new mount in late 2021, resulting in two identical 8-UT telescopes as pictured in Figure~\ref{fig:photo2}. Throughout 2021-22 these new telescopes have been undergoing commissioning, hampered by the ongoing COVID-19 pandemic and the eruption of the Cumbre Vieja volcano on La Palma from September--December 2021.

\subsection{GOTO-South (2022+)}
\label{sec:phase3}

In addition to the two telescopes forming the GOTO-North node on La Palma, it was always anticipated that a second node would be constructed in the southern hemisphere to open up the entire sky when searching for gravitational-wave counterparts. Furthermore, a second site $180^{\circ}$ removed in longitude from La Palma would ensure close to 24-hour capability to respond to transient alerts due to the alternating dark times. Based on these factors, as well as the sensitivity maps of the LIGO-Virgo gravitational-wave detectors \cite{GWprospects}, Australia was the obvious location to host the GOTO-South node, with eastern Australia being near-antipodal to the Canary Islands.

After considering multiple possible sites in the southern hemisphere, it was decided to locate GOTO-South at Siding Spring Observatory in New South Wales, Australia. Siding Spring was judged to provide the best observing conditions as well as the benefit of an established observatory with existing infrastructure and facilities. As of mid-2022 excavation has begun on the concrete platform on which the domes will be constructed. Commissioning of the first GOTO telescope in Australia is expected to begin in late-2022, with the fourth following in 2023.

The next major milestone for the project will be the start of the start of the O4 LIGO-Virgo-KAGRA observing run, which at the time of writing is scheduled for early 2023. By then GOTO aims to have at least three fully-operational telescopes, two in the north, one in the south, with the second southern telescope coming online during the observing run. 

\newpage

\section{Conclusion}
\label{sec:conclusion}

We have presented the GOTO project, which aims to build a network of wide-field robotic telescopes across the globe dedicated to detecting electromagnetic counterparts to gravitational-wave events. The project is currently in a critical phase of deployment, with the ongoing commissioning of two new telescopes at the GOTO-North site in La Palma alongside construction beginning of the southern site in Australia. Once fully deployed each telescope will be networked to form a robotic, multi-site observatory, which will survey the entire visible sky every two nights and enable rapid follow-up detections of transient sources. When fully operational the GOTO network will be a powerful transient discovery engine, able to respond rapidly to transient event alerts while also providing a high-cadence survey for numerous secondary science projects.

\section*{Acknowledgements}

The Gravitational-wave Optical Transient Observer (GOTO) project acknowledges the support of the Monash– Warwick Alliance; the University of Warwick; Monash University; the University of Sheffield; the University of Leicester; Armagh Observatory \& Planetarium; the National Astronomical Research Institute of Thailand (NARIT); the Instituto de Astrofísica de Canarias (IAC); the University of Portsmouth; the University of Turku; the University of Manchester and the UK Science and Technology Facilities Council (STFC; grant numbers ST/T007184/1, ST/T003103/1, and ST/T000406/1).

\bibliography{report}
\bibliographystyle{spiebib_shortlist}


\end{document}